\documentclass[11pt]{article}

\usepackage{fullpage}
\usepackage{amssymb}
\usepackage{graphics}
\usepackage{epsfig}
\usepackage{psfrag}
\usepackage{graphicx, color, float}
\usepackage{latexsym}
\usepackage{amsmath}

\newtheorem{theorem}{\bf Theorem}[section]
\newtheorem{corollary}[theorem]{\bf Corollary}

\newtheorem{proposition}[theorem]{\bf Proposition}

\newcommand{\Cost} {{\rm Cost}}
\newcommand{\OPT} {{\rm OPT}}
\newcommand{\proof} {{\it Proof: }}
\newcommand{\qed} {\hfill$\Box$}
\newcommand{\calP} {{\cal P}}
\newcommand{\calR} {{\cal R}}
\newcommand{\Chlebik} {Chleb\'{\i}k}
\newcommand{\Chlebikova} {Chleb\'{\i}kov{\'a}}

\title{
On the Complexity of the $k$-Anonymization Problem
}
\author{Venkatesan T. Chakaravarthy, Vinayaka Pandit, Yogish Sabharwal}
\date{
IBM Research - India, New Delhi and Bengaluru.\\
{\it \{vechakra, pvinayak, ysabharwal\}@in.ibm.com}
}

\begin{document}
\maketitle

\begin{abstract}
We study the problem of anonymizing tables containing personal information
before releasing them for public use. One of the formulations considered in this
context is the $k$-anonymization problem: given a table, suppress a minimum
number of cells so that in the transformed table, each row is identical to
atleast $k-1$
other rows. The problem is known to be NP-hard and MAXSNP-hard; but in the
known reductions, the number of columns in the constructed tables is arbitrarily
large. However, in practical settings the number of columns is much smaller.
So, we study the complexity of the practical setting in
which the number of columns $m$ is small. We show that the problem is NP-hard,
even when the number of columns $m$ is a constant ($m=3$). We also prove
MAXSNP-hardness for this restricted version and derive that the problem cannot
be approximated within a factor of $\frac{6238}{6237}$. Our reduction uses
alphabets $\Sigma$ of arbitrarily large size. A natural question is whether the
problem remains NP-hard when both $m$ and $|\Sigma|$ are small. We prove that
the $k$-anonymization problem is in $P$ when both $m$ and $|\Sigma|$ are constants.
\end{abstract}

\section{Introduction}
\label{sec:intro}
Various organization such as hospitals and insurance companies collect massive
amount of personal data. These need to be released publicly for the purpose of
scientific data mining; for instance, data collected by hospitals could be mined
to infer epidemics. However, a major risk in releasing personal data is that
they can be used to infer sensitive information about individuals. A natural
idea for protecting privacy is to remove obvious personal identifiers such as
social security number, name and driving license number. However,
Sweeney~\cite{Swe02} showed that such a deidentified database can be joined with
other publicly available databases (such as voter lists) to reidentify
individuals. For instance, she showed that 87\% of the population of the United
States can be uniquely identified on the basis of gender, date of birth and
zipcode. In the literature, such an identity leaking attribute combination is
called a quasi-identifier. It is important to recognize quasi-identifiers and
apply protective measures to eliminate the risk of identity disclosure via join
attacks. Samaratti and Sweeyney~\cite{Sam-Swe, Swe02} introduced the notion of
$k$-anonymity, which aims to preserve privacy either by suppressing or
generalizing some of the sensitive data values.

%Subsequently, $k$-anonymity and its variations have been well studied 
%(see~\cite{cluster-pods, Lefevre1, Lefevre2, l-diversity}).

In this paper, we consider the basic $k$-anonymity problem with only suppression allowed.
Suppose we have a table with $n$ rows and $m$ columns. In order to achieve anonymity, one is allowed
to suppress the entries of the table so that in the modified table, every row is identical to at least
$k-1$ other rows. The goal is to minimize the number of cells suppressed.
This is called the {\em $k$-anonymization} problem.
The motivation for the problem formulation are twofold: (i) any join attack would return groups of at least $k$ rows,
thus preserving privacy with a parameter of $k$; (ii) lesser the number of entries suppressed,
better is the value of the modified table for data mining.

{\bf Example: }We now illustrate the problem definition with an example. 
An example input table and its anonymized output, for $k=2$, are shown in Figure~\ref{fig:example}.
The number of rows is $n=4$ and number of columns is $m=3$.
The suppressed cells are shown by ``$*$''. 
We see that in the anonymized output table, the first and the third rows are identical, 
and the second and the fourth rows are identical. Thus the table on the right is $2$-anonymized.
The cost of this anonymization is $4$, since $4$ cells are suppressed. This is an optimal solution.
\begin{figure}
\begin{center}

\begin{tabular}{ccccc}

\begin{tabular}{|c|c|c|}
\hline
$x$ & $a$ & $b$ \\ 
\hline
$z$ & $c$ & $d$ \\
\hline
$y$ & $a$ & $b$ \\
\hline
$z$ & $c$ & $e$ \\
\hline
\end{tabular}

&
&
&
&

\begin{tabular}{|c|c|c|}
\hline
$*$ & $a$ & $b$ \\ 
\hline
$z$ & $c$ & $*$ \\
\hline
$*$ & $a$ & $b$ \\
\hline
$z$ & $c$ & $*$ \\
\hline
\end{tabular}

\\
&&&&
\\
Original table
&
&
&
&
$2$-Anonymized table
\end{tabular}
\end{center}
\caption{An Example}
\label{fig:example}
\end{figure}

\noindent
{\bf Known and New Results: }

Meyerson and Williams~\cite{MW} proved the NP-hardness of the $k$-anonymization problem.
Aggarwal et al.~\cite{ICDT} improved the result by showing that the problem remains NP-hard
even when the alphabet $\Sigma$ from which the symbols of the table are drawn is fixed to be ternary.
Bonizzoni et al~\cite{APX-hard} proved  MAXSNP-hardness (and NP-hardness) even when the alphabet is binary.
The value of the privacy parameter $k$ is a fixed constant in all the above results ($k=3$). 
On the algorithmic front, Meyerson and Williams gave a $O(k\log k)$-approximation algorithm.
This was improved by Aggarwal et al.~\cite{ICDT}, who devised a $O(k)$-approximation algorithm. 
Park and Shim~\cite{SIGMOD} presented an approximation algorithm with a ratio of $O(\log{k})$; 
however, we observe that the running time of their algorithm is exponential in the number of columns $m$
(but, polynomial in the number of rows $n$).

We make the following observations regarding the previously known results.
Firstly, the known NP-hardness reductions produce tables in which the number of
columns is arbitrarily large. This is not satisfactory as the number of columns
in practical settings is not large. Secondly, the algorithm of Park and
Shim~\cite{SIGMOD} is a polynomial time $O(\log k)$-approximation algorithm when
the number of columns $m$ is small ($m=O(\log n)$). These observations raise a
natural question: Does the $k$-anonymization problem remain NP-Complete even
when the number of columns $m$ is small ($\log n$ or a constant)? 
We show that the $k$-anonymization problem remains NP-hard, even
when the number columns $m$ is fixed to be a constant ($m=3$). In fact, we also
show that the above restricted version is MAXSNP-hard, thus ruling out
polynomial time approximation schemes. We also derive that the problem cannot be
approximated within a factor of $\frac{6238}{6237}$. Even though our
inapproximability bound is mild, it is the first explicit inapproximability
bound proved for the $k$-anonymization problem. All our hardness results hold
even when the privacy parameter $k$ is a constant ($k=7$).

%As can be seen, there exists a large gap between the known hardness and the approximation algorithm guarantees.
%On the hardness side, only MAXSNP-hardness~\cite{APX-hard} is known for the problem,
%which means that the problem cannot be approximated within a factor of $(1+\epsilon)$, for some $\epsilon > 0$.
%Whereas on the algorithmic side, the best known approximation guarantee for a polynomial time algorithm is 
%$O(k)$~\cite{ICDT}. Aggarwal et al.~\cite{ICDT} used a natural graph theoretic framework to devise the
%$O(k)$-approximation algorithm and also showed that any polynomial time algorithm that uses this framework cannot
%achieve a factor better than $O(k)$. Thus, breaking the $O(k)$ barrier seems to be a hard problem,
%requiring highly sophisticated algorithms. 
%Alternatively, one can study restricted versions of the problem under practical settings. 
%Park and Shim~\cite{SIGMOD} provide a $O(\log k)$ polynomial time approximation 
%algorithm, when the number of columns is $m=O(\log n)$. Since $m$ is small in practice, studying the 
%the above restriction is useful. 
%Incidentally, the known hardness results~\cite{MW, ICDT, APX-hard}
%do not hold for this practical setting, since the number of columns of the tables constructed in these reductions
%is arbitrarily large and is typically, more than the number of rows. 
%Therefore, an interesting open problem here is whether the problem can be
%solved optimally in polynomial time when $m$ is small (i.e., $m=O(\log n)$).

As we noted, the previous constructions ensured that the alphabet size is a fixed constant;
but, in our constructions, the alphabet size is not a fixed constant, but it is arbitrarily large.
However, this is not a serious issue; in most settings, tables have large number of unique entries
(for example, a zipcode column takes a large number of distinct values).
In the wake of previous results and our results mentioned above, 
a natural question is whether the problem is NP-hard when both the number of columns
$m$ and the alphabet size $|\Sigma|$ are small. 
We show that the problem can be solved optimally in polynomial time when both $m$ and $|\Sigma|$ are fixed constants.

\iffalse
Below, we summarize the new results proved in this paper.
\begin{itemize}
\item
The $k$-anonymization problem is NP-hard even when the number of columns $m$ and the privacy
parameter $k$ are constants ($m=3$, $k=7$). 
We show that the restricted version is MAXSNP-hard and cannot be approximated within a factor
of $\frac{6238}{6237}$.
\item
When both the number of columns and the alphabet size are fixed constants, the problem can be solved
optimally in polynomial time. The result is true, even when the privacy parameter $k$ is arbitrarily large.
\end{itemize}

The known and new results put together provide a complete picture on the complexity
of the problem for the four cases of the number of columns $m$ and the alphabet size $\Sigma$ 
being fixed constants or arbitrarily large. This is shown in Figure~\ref{fig:summary}.

\begin{figure}
\begin{center}
\begin{tabular}{||c|p{6cm}|p{6cm}||}
\hline
\hline
& & \\
& \ \ \ \ $\mathbf{|\Sigma|}$\bf{=constant} & \ \ \ \ $\mathbf{|\Sigma|}$\bf{=arbitrary}\\
& & \\
\hline
& & \\
$\mathbf{m}$\bf{=constant} & Solvable in P (even when $k$ is arbitrary) [This paper] & NP-hard (even when $k$ is a constant) [This paper]\\
& & \\
\hline
& & \\
$\mathbf{m}$\bf{=arbitrary} & NP-hard (even when $k$ is a constant)~\cite{ICDT} & NP-hard (even when $k$ is a constant)~\cite{ICDT}\\
& & \\
\hline
\hline
\end{tabular}
\end{center}
\caption{Summary of known and new results}
\label{fig:summary}
\end{figure}
\fi

\section{Problem Definition}
%In this section, we formally define the $k$-anonymization problem. We also develop some useful notation
%and present a few remarks. 

The input to the $k$-anonymization problem is a $n\times m$ table $T$ having $n$
rows and $m$ columns, with symbols of the table drawn from an alphabet $\Sigma$.
The input also includes a {\em privacy parameter} $k$. A feasible solution
$\sigma$ transforms the given table $T$ to a new table $T'$ by suppressing some
of the cells of $T$; namely, it replaces some of the cells of $T$ with ``$*$''.
In the transformed table $T'$, for any row $t$, there should exist $k-1$ other
rows that are identical to $t$. The cost of the solution, denoted
$\Cost(\sigma)$, is the number of suppressed cells. The goal is to find a
solution having the minimum cost. Consider a solution $\sigma$. For a row $t$,
we denote by $\Cost(t)$ the number of suppressed cells in $t$ and say that $t$
pays this cost. Thus, $\Cost(\sigma)$ is the sum of costs paid by all the rows.

There is an equivalent way to view a solution in terms of partitioning the given table.
Consider a subset of rows $S$. We say that a column is {\em good} with respect to $S$, if all the rows
in $S$ take identical values on the column. A column is said to be {\em bad}, if it is not good;
meaning, some two rows in $S$ have different values on the given column. 
Denote by $a(S)$ the number of bad columns in $S$. Then, our goal is to a find a partition of rows
$\Pi={S_1, S_2, \ldots, S_{\ell}}$ such that each set $S_i$ is of size $|S_i|\geq k$.
Each row $t$ in $S_i$ pays a cost of $a(S_i)$. The total cost of the solution is the
sum of costs paid by all rows. Equivalently, the cost of the solution is given
by $\sum_{i=1}^{\ell} |S_i|\cdot a(S_i)$.

We shall interchangeably use either of the two descriptions in our discussions.

\section{Hardness Results with Three Columns}
In this section, we present results on the complexity of $k$-anonymization
problem when both the number of columns and the privacy parameter are constants. 
\subsection{NP-Hardness}

\begin{theorem}
\label{thm:AAA}
The $k$-anonymization problem is NP-hard 
even when the number of columns $m$ is 3 and the privacy parameter is fixed as $k=7$.
\end{theorem}
\proof
We give a reduction from the vertex cover problem on 3-regular graphs, which is known to be
NP-hard (see~\cite{GJ}). Recall that a vertex cover of a graph refers to a subset of vertices
such that each edge has at least one endpoint in the subset and that a graph is said to be 3-regular,
if every vertex has degree exactly 3.

Let $G=(V,E)$ be the input 3-regular graph having $r$ vertices.
The alphabet of the output table is as follows: 
For each vertex $u \in V$, we add a symbol $u$. Next, we have additional symbols
`$0$' and `$Z$'. Further, we need a number of {\em special} symbols. A special symbol
appears only once in the whole of the table. The exposition becomes somewhat clumsy,
if we explicitly introduce these special symbols. Instead, we use the generic symbol
`$?$' to mean the special symbols. The symbol `$?$' is not a single symbol,
but a general placeholder to mean a special symbol. We maintain a running list
of special symbols (say $s_1, s_2, \ldots$) and whenever a new row containing `$?$' is 
added to the table, we actually get a new symbol from the list and replace `$?$'
by the new symbol. For instance, suppose $\langle ?, u, u\rangle$ and $\langle ?, v, ?\rangle$
are the first two rows added to the table. Then, the actual rows added are $\langle s_1, u, u\rangle$
and $\langle s_2, u, s_3\rangle$. With the above discussion in mind, notice that
two different instances of `$?$' do not match with each other.

The output table $T$ is constructed as follows. 
\begin{enumerate}
\item
   For each vertex $u\in V$, add the following 20 rows. These are said to be rows corresponding to $u$.
   \begin{enumerate}
   \item
   Add the following row six times: $\langle 0, u, u\rangle$.
   \label{type:AAA}
   \item
   Add a row $\langle ?, u, u\rangle$. It is called the {\em critical} row of $u$
   and it plays a vital role in the construction.
   \label{type:BBB}
   \item 
   Add seven rows: $\langle ?, u, ?\rangle$.
   \label{type:CCC}
   \item 
   Add the following row 3 times: $\langle 0, u, Z\rangle$.
   \label{type:DDD}
   \item
   Add the following row 3 times: $\langle 0, Z, u\rangle$.
   \label{type:EEE}
   \end{enumerate}
\item
   For each edge $(x, y)$, add two rows $\langle 0, x, y\rangle$ and $\langle 0, y, x\rangle$.
   These are called {\em edge rows}.
\item 
   Add the following two sets of {\em dummy rows}:
   \begin{enumerate}
   \item
   Add seven rows as below: $\langle 0, ?, Z\rangle$. 
   \label{type:DDD-A}
   \item
   Add seven rows as below: $\langle 0, Z, ?\rangle$. 
   \label{type:DDD-B}
   \end{enumerate}
\end{enumerate}
This completes the construction of the table. The privacy parameter is set as $k=7$.

Consider any $k$-anonymization solution to the constructed table. 
For any row of the table,  we can derive a lowerbound on the cost paid by the row;
we refer to the lowerbound as the {\em base cost}. The base costs are derived as follows,
for the various types of rows.
Consider any vertex $u\in V$. First consider rows of type~\ref{type:AAA}.
These rows are of the form $\langle 0, u, u\rangle$ and there are exactly six of them.
Since $k=7$, these rows must be participating in a cluster having a different row. 
Hence, each of these rows must pay a cost of at least 1. 
We set the base cost for each of these rows to be 1. 
Now, consider the critical row of type~\ref{type:BBB}.
This row is of the form $\langle ?, u, u\rangle$ and 
it must pay a base cost of 1, since it involves a special symbol.
The base cost of the critical row is deemed to be 1.
A type~\ref{type:CCC} row (of the form $\langle ?, u, ?\rangle$) must pay cost of at least two,
since it has two special symbols. The base cost of such a row is deemed to be 2.
By similar arguments, we see that any other type of row must pay a base cost of 1.
To summarize, every row of type~\ref{type:CCC} (of the form $\langle ?, u, ?\rangle$)
pays a base cost of 2, whereas any row of any other type pays a base cost of 1.

For each vertex $u$, the total base cost across the 20 rows can be calculated as follows:
(i) The six (type~\ref{type:AAA}) rows of the form $\langle 0, u, u\rangle$  pay a cost of $6$ in total;
(ii) The critical row (of type~\ref{type:BBB}) pays a cost of $1$;
(iii) The seven (type~\ref{type:CCC}) rows of the form $\langle ?, u, ?\rangle$ 
pay a cost of $2$ each, totaling $14$;
(iv) The three (type~\ref{type:DDD}) rows of the form $\langle 0, u, Z\rangle$ pay cost of $3$ in total;
(v) The three (type~\ref{type:EEE}) rows of the form $\langle 0, Z, u \rangle$ pay cost of $3$ in total.
Thus, the total base cost for each vertex $u$ is $27$.
Then, each edge has a base cost of 2, coming from the two rows corresponding to it.
The two blocks dummy rows (of type~\ref{type:DDD-A} and type~\ref{type:DDD-B}) contribute a base cost of $7$ each,
summing up to $14$.
Thus, the {\em aggregated base cost} is $ABC = 27r + 2|E|+14$.
For any row, the difference between the actual cost paid and the base cost is denoted as {\em extra cost}.
Similarly, the total extra cost is the sum of extra costs over all the rows.
Notice that the cost of the solution is the sum of $ABC$ and the total extra cost.

We claim that the given graph has a vertex cover of size $\leq t$, if and only if there
exists a $k$-anonymization solution with an extra cost $\leq t$.
It would follow that the graph has a vertex cover of size $\leq t$, if and only if there
exists a $k$-anonymization solution of cost $\leq ABC+t$.
This would prove the required NP-hardness. We next proceed to prove the above claim.
We split the proof into two parts.

First, we shall argue that if the given graph has a vertex cover of size $\leq t$, then there exists
a $k$-anonymization solution with extra cost $\leq t$.
Suppose $C$ is a vertex cover of size $\leq t$.
We shall construct a $k$-anonymization solution $\sigma$ in which the critical rows corresponding
to the vertices in the cover $C$ pay an extra cost of 1 and every other row pays no extra cost.

For each edge $(x, y)$, if $x\in C$, then {\em attach} the edge to $x$, else attach it to $y$.
(If both the endpoints of the edge are in the cover, the edge can be attached arbitrarily to any one of the two vertices).
Without loss of generality, assume that each vertex in the cover has at least one edge attached to it.
Otherwise, the vertex can be safely removed from $C$, yielding a smaller cover.

\begin{figure}
\begin{center}
\begin{tabular}{|c|c|}
    \hline
    \psfig{figure=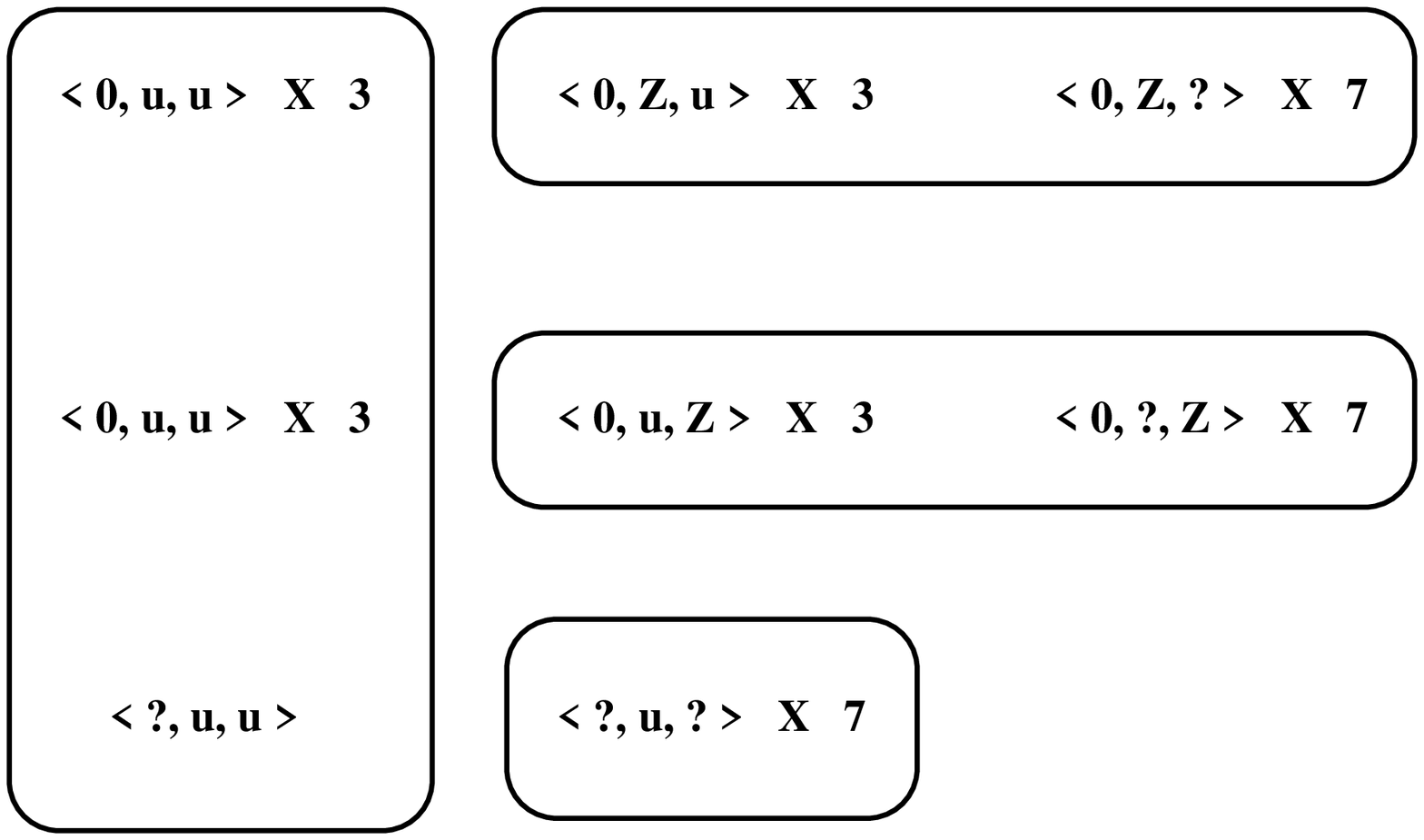, width=0.4\columnwidth} &
    \psfig{figure=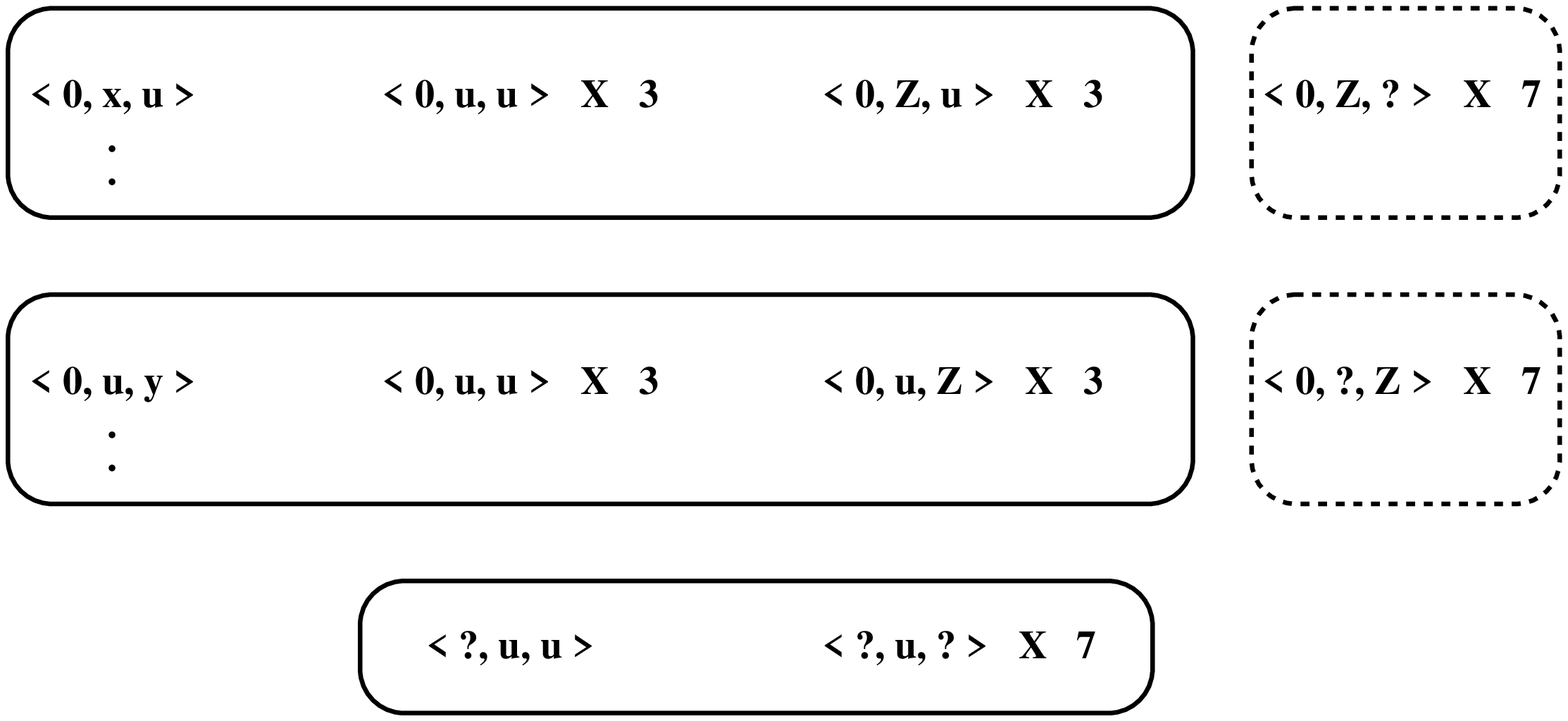, width=0.5\columnwidth} \\
    Case: $u\not\in C$ & Case: $u\in C$\\
    \hline
\end{tabular}
\end{center}
\caption{Construction of $\sigma$ from cover $C$}
\label{fig:hard}
\end{figure}

Form a $k$-anonymization solution $\sigma$ as follows. See Figure~\ref{fig:hard}
for an illustration.
\begin{itemize}
\item
   Form two clusters combining the dummy rows.
   \begin{itemize}
   \item
   Form a cluster by combining the seven (type~\ref{type:DDD-A}) dummy rows of the form $\langle 0, ?, Z\rangle$; 
   call this cluster $D_1$.
   \item
   Form a cluster by combining the seven (type~\ref{type:DDD-B}) dummy rows of the form $\langle 0, Z, ?\rangle$; 
   call this cluster $D_2$.
   \end{itemize}
\item
Consider each vertex $u$ not in the cover $C$ (i.e., $u\not\in C$).
   \begin{itemize}
   \item
   Form a cluster by adding the six (type~\ref{type:AAA}) rows of the form $\langle 0, u, u\rangle$ and 
   the critical row $\langle ?, u, u\rangle$.
   Each row in the cluster pays a cost of 1, and hence the extra cost is 0 for all these rows.
   \item
   Form a cluster by adding the seven (type~\ref{type:CCC}) rows of the form $\langle ?, u, ?\rangle$.
   Each row in the cluster pays a cost of 2, and hence the extra cost is 0 for all these rows.
   \item
   Add the three (type~\ref{type:DDD}) rows of the form $\langle 0, u, Z\rangle$ to $D_1$.
   Add the three (type~\ref{type:EEE}) rows of the form $\langle 0, Z, u\rangle$ to $D_2$.
   Thus, each row in $D_1$ and $D_2$ pays a cost of 1, and hence their extra costs are 0.
   \end{itemize}
\item
Consider each vertex $u$ in the cover $C$ (i.e.,  $u\in C$).
   \begin{itemize}
   \item
   Form a cluster $A_u$ by adding three of the (type~\ref{type:AAA}) rows of the form 
   $\langle 0, u, u\rangle$
   \item
   Form a cluster $B_u$ by adding the remaining three (type~\ref{type:AAA}) rows of the form 
   $\langle 0, u, u\rangle$
   \item
   Consider each edge attached to $u$, say $(u, x)$ for some $x\in V$.
   Add the edge row $\langle 0, u, x\rangle$ to $A_u$
   and add the edge row $\langle 0, x, u\rangle$ to $B_u$.
   \item
   Add the three (type~\ref{type:DDD}) rows of the form $\langle 0, u, Z\rangle$ to $A_u$
   and add the three (type~\ref{type:EEE}) rows of the form $\langle 0, Z, u\rangle$ to $B_u$.
   Notice that both $A_u$ and $B_u$ have at least seven rows each, since each vertex has at least
   one edge attached to it.
   Every row in these two clusters pays a cost of $1$ (thus, the extra cost paid by these rows is 0).
   \item 
   Form a cluster by adding the seven (type~\ref{type:CCC}) rows of the form $\langle ?, u, ?\rangle$.
   Add the critical row $\langle ?, u, u\rangle$ to this cluster.
   Notice that the seven rows each pay a cost of $2$, and hence their extra cost is $0$.
   {\em The critical row pays a cost of $2$, and hence, its extra cost is $1$}.
   \end{itemize}
\end{itemize}
Observe that all the rows of the table have been assigned to some cluster and each cluster has size at least $7$.
From the above discussion, we see that the only rows having non-zero extra cost are the critical rows
corresponding to the vertices in the cover $C$ and they pay an extra cost of $1$ each. 
We conclude that the total extra cost is $|C|$.
We have proved the following claim:

{\sc Claim 1: }
If the given graph has a vertex cover of size $\leq t$, then there exists
a $k$-anonymization solution with extra cost $\leq t$.

We next proceed to prove the reverse direction: 
if there exists a $k$-anonymization solution $\sigma$ of extra cost $\leq t$,
then there exists a vertex cover of size $\leq t$.
Consider such a solution $\sigma$. We first make the following claim.

{\sc Claim 2: }
Consider a vertex $u$. Suppose the critical row $\langle ?, u, u\rangle$
pays an extra cost of 0. Then, the only cluster in which it can participate is the one
obtained by combining the critical row with the six (type~\ref{type:AAA}) rows of the 
form $\langle 0, u, u\rangle$.\\
\proof
Clearly, the critical row must pay a cost of $1$, since it has a special symbol.
If it pays no extra cost, then the rows it is combined with should have the symbol '$u$'
in their second and third columns. There are exactly six such rows available and these are the (type~\ref{type:AAA})
rows of the form $\langle 0, u, u\rangle$.
\qed

We say that a vertex is {\em perfect}, if all the 20 rows corresponding to it pay an extra cost of 0.
A vertex is said to be {\em imperfect}, if at least one of its 20 rows pay an extra cost of at least 1.

{\sc Claim 3: }Consider an edge $(x, y)$. If both $x$ and $y$ are perfect, then 
at least one of the two edge rows corresponding to the edge pays a cost of at least 2.
\\
\proof
Consider the edge row $\langle 0, x, y\rangle$, corresponding to the given edge.
Let $S$ be the cluster to which this row belongs.
Since $k=7$, we have $|S|\geq 7$.
Recall that a column is said to be good with respect to $S$, 
if the rows of $S$ have identical values on the column;
a column is said to be bad with respect to $S$, otherwise.
We shall argue that at least two of the three columns are bad with respect to $S$. 
Let us consider the three possible choices for two-column subsets out of the three columns.
\begin{itemize}
\item
Clearly, both the second and the third columns cannot be good with respect to $S$,
since there are no other rows that contain $x$ in their second column and $y$ in their third column.
\item
Next, we argue that both the first and the second column cannot be good with respect to $S$.
Since, both $x$ and $y$ are perfect, their critical rows do not pay any extra cost.
By Claim 2, the six (type~\ref{type:AAA}) rows of the form $\langle 0, x, x\rangle$ 
have gone to some cluster other than $S$.
Similarly, the six (type~\ref{type:AAA}) rows of the form $\langle 0, y, y\rangle$ have also gone to 
some other cluster. These rows cannot be part of $S$.
Now, since the graph is 3-regular, there are only two other edge rows that have `$0$' in their first column 
and `$x$' in their second column; these correspond to the two other edges incident on $x$. 
There are three other rows (corresponding to the vertex $x$ and of type~\ref{type:DDD})
that have `$0$' in their first column and `$x$' in their second column.
Thus, totally there are only 5 other rows that that have the above property.
Since $|S|\geq 7$, it follows that both the first and the second column cannot be good with respect to $S$.
\item
A similar argument shows that there are only 5 other rows that have `$0$' in their first column and 
`$y$' in their third column. 
This means that both the first column and the third column cannot be good in $S$.
\end{itemize}
We conclude at least two of the three columns are bad respect to $S$.
Thus, the concerned edge row $\langle 0, x, y\rangle$ must pay a cost of at least 2.
\qed

Let $V'$ be the set of all imperfect vertices. Let $E'$ be the set edges
whose both endpoints are perfect. Each imperfect vertex (by definition) contributes at least 
1 to the extra cost. By Claim 3, each edge in $E'$ pays an extra cost of at least 1.
Therefore,
\[
\mbox{total extra cost of $\sigma$} \geq |V'| + |E'|.
\]

Construct a vertex cover $C$ as follows. Add every imperfect vertex to $C$.
For each edge in $E'$, add one of its endpoints (arbitrarily) to $C$. 
Clearly, $C$ is a vertex cover. So,
\[
|C| \leq |V'|+|E'| \leq \mbox{extra cost of $\sigma$} 
\]

We have proved the following claim:

{\sc Claim 4: }
If there exists a $k$-anonymization solution $\sigma$ of extra cost $\leq t$,
then there exists a vertex cover of size $\leq t$.
\qed

We observed that the cost of a $k$-anonymous solution is the sum of $ABC$ and the extra cost of the solution.
Now, by combining Claim 1 and Claim 4, we get the following:
there exists a vertex cover of size $\leq t$, if and only if there exists a $k$-anonymization solution
of cost $\leq ABC+t$. This completes the NP-hardness proof.
\qed

It is easy to show that our reduction is an $L$-reduction (see~\cite{Papa} for a
discussion on $L$-reductions). As the vertex cover problem
on 3-regular graphs is MAXSNP-hard~\cite{Alimonti}, it follows that,
\begin{theorem}
\label{thm:BBB}
The $k$-anonymization problem is MAXSNP-hard, even when the number of columns in
3 and the privacy parameter
is fixed as $k=7$.
\end{theorem}

Moreover, 
{\Chlebik} and {\Chlebikova}~\cite{Chlebik} showed that the vertex cover problem on 3-regular graphs
cannot be approximated within a factor of $\frac{100}{99}$. Now, taking the
parameters of the $L$-reduction of our construction, and based on the result of
{\Chlebik} and {\Chlebikova}, we can show that,
\begin{corollary}
The $k$-anonymization problem cannot be approximated within a factor of $\frac{6238}{6237}$, 
even when the number of columns in 3 and the privacy parameter is fixed as $k=7$.
\end{corollary}

\section{Special case: $m$ and $|\Sigma|$ are constants}
As our NP-hardness reduction utilizes alphabets of
arbitrarily large size, a natural question is whether the problem remains
NP-hard when both the number of columns and the alphabet size are fixed
constants. Here, we show that this case can be solved optimally in polynomial
time.

In the problem definition, let $m$, the number of columns, to be a constant and
let the size of $\Sigma$ be a constant $s$ and let $k$ be the privacy
parameter. By a {row pattern}, we mean a vector over $m$ columns whose entries
belong to $\Sigma$. Let $\calR$ denote the set of all row patterns; $|\calR| =
|\Sigma|^{m}$ is a constant. By an {\em anonymization pattern}, we mean a
vector over $m$ columns whose each entry is either a symbol from $\Sigma$ or the
suppression symbol `$*$'. Let $\calP$ denote the set of all anonymization
patterns; $|\calP| = (|\Sigma|+1)^{m+1}$ is a constant . We say that a row pattern $t$
{\em matches} an anonymization pattern $p$, if $p$ and $t$ agree on all columns,
except the columns suppressed in $p$. We use ``$t\sim p$" as a shorthand to mean
that $t$ matches $p$. Consider the optimal $k$-anonymization solution
$\sigma^*$. For each row pattern $t$, the solution $\sigma^*$ chooses an anonymization
pattern $p$ matching $t$ and applies $p$ to $t$. If a pattern $p\in \calP$ is
applied to a row pattern $t$, we say that $t$ is {\em attached} to $p$. The solution
satisfies the property that, for each anonymization pattern $p \in \calP$,
number of row patterns attached to it is either zero or at least $k$. If no row pattern is
attached to $p$, then we say that $p$ is {\em closed}; on the other hand, if at
least $k$ row pattern are attached to $p$, we say that $p$ is {\em open}. Thus, the
optimal solution $\sigma^*$ opens up some subset of patterns from $\calP$. Of
course, we do not know which patterns are open and which are closed. But, we
can guess the set of open patterns by iterating over all possible subsets of
$\calP$. For each subset $P\subseteq \calP$, our goal is to compute the optimal
solution whose set of open patterns is exactly equal to $P$. The number of such
subsets is $2^{|\calP|}$, which is a constant since $|\calP|$ is a constant.
Then, we take the minimum of the over these solutions.

Consider a subset of patterns $P$. Our goal is to find the optimal solution in
which the set of open patterns is exactly equal to $P$.
Notice that there may not exist any feasible solution for the subset $P$;
we also need to determine, if this is the case. This can be formulated as the
following integer linear program. For each row pattern $t \in \calR$, 
$s(t)$ denotes the number of copies (i.e., tuples) of the row pattern in the input
table ($s(t) = 0$ if the row pattern $t$ does not occur in the table).
For each pair $(p \in \calP,t\in \calR)$ such that the row pattern $t$
matches the pattern $p$, we introduce an integer variable $x_{p,t}$. This
variable captures the number of copies of the row pattern $t$ attached to the
anonymization pattern $p$. For a
pattern $p$, let $\Cost(p)$ denote that number of suppressed cells in $p$; this
is the cost each copy of a row pattern $t$ would pay, if $t$ is attached to $p$.
\begin{eqnarray}
\notag
\min \sum_{(p,t):t\sim p} \Cost(p) x_{p,t} & &\\
\notag
\mbox{subject to:} &\\
\sum_{t:t\sim p} x_{p,t} \geq k & & \mbox{for all } p\in P\\
\label{eqn:AAA}
\sum_{p:t\sim p} x_{p,t} = s(t) & & \mbox{for all } t\in \calR\\
\label{eqn:BBB}
x_{p, t}\in \mathbb{N}_0 & & \mbox{for all } (p,t):t\sim p
\end{eqnarray}

Note that this integer linear program has a constant number of variable as the
number of $x_{p,t}$ variables is bounded by $|\calP|\cdot|\calR| \leq
m^{2|\Sigma|+1}$. By the famous result of Lenstra~\cite{Lenstra}, an integer linear program on
constant number of variables can be solved in polynomial time.

This approach, when applied to the practical case of $m=O(\log n)$ and
$|\Sigma|$ being arbitrarily large, leads to a variant of facility location problem.
The patterns ($n2^m=n^{O(1)}$ in number) can be viewed as facilities with a connection cost equal to the number of suppressed cells.
The rows can be viewed as clients who can be serviced by any pattern that they match to.
The goal is to open a subset of the facilities and attach the clients to the facilities such
that every open facility has at least $k$ clients attached to it.
Objective is to minimize the total connection cost of all the clients. Note that
the distances here are non-metric. No approximation algorithms are known for
this variant. Designing approximation algorithms for
this facility location problem that can in turn yield approximation algorithm
for the above case of anonymization problem would be interesting.
\vspace*{-0.3cm}

\section{Open Problems}
For the general $k$-anonymization problem, the best known approximation
algorithm, due to Aggarwal et al.~\cite{ICDT}, achieves a ratio of $O(k)$. Their
algorithm is based on a natural graph theoretic framework. They showed that any
poly-time algorithm that uses their framework cannot achieve a factor better
than $O(k)$. Breaking the $O(k)$-approximation barrier seems to be a challenging
open problem. Improving the $O(\log k)$ approximation ratio, due to Park and
Shim~\cite{SIGMOD}, for the practical special case when $m=O(\log n)$ is an
interesting open problem. For the case where $m$ is constant, a trivial constant
factor approximation algorithm exists: suppressing all cells yields an $O(m)$
approximation ratio. However, it is challenging to design an algorithm that, for
all constants $m$, guarantees a fixed constant approximation ratio (say, $2$);
notice that such an algorithm is allowed to run in time $2^{2^m}$. Getting a
hardness of approximation better than $\frac{6238}{6237}$ would be of interest.
\bibliographystyle{plain}
\bibliography{anon}

\begin{thebibliography}{10}

\bibitem{ICDT}
G.~Aggarwal, T.~Feder, K.~Kenthapadi, R.~Motwani, R.~Panigrahy, D.~Thomas, and
  A.~Zhu.
\newblock Anonymizing tables.
\newblock In {\em ICDT}, 2005.

\bibitem{Alimonti}
P.~Alimonti and V.~Kann.
\newblock Hardness of approximating problems on cubic graphs.
\newblock In {\em 3rd Italian Conference on Algorithms and Complexity}, 1997.

\bibitem{APX-hard}
P.~Bonizzoni, G.~Vedova, and R.~Dondi.
\newblock Anonymizing binary tables is apx-hard.
\newblock {\em CoRR}, abs/0707.0421, 2007.

\bibitem{Chlebik}
M.~Chleb\'{\i}k and J.~Chleb\'{\i}kov{\'a}.
\newblock Complexity of approximating bounded variants of optimization
  problems.
\newblock {\em Theoretical Computer Science}, 354(3):320--338, 2006.

\bibitem{GJ}
M.~Garey and D.~Johnson.
\newblock {\em Computers and Intractability: A Guide to the Theory of
  NP-Completeness}.
\newblock Freeman, 1979.

\bibitem{Lenstra}
H.Lenstra.
\newblock Integer programming with a fixed number of variables.
\newblock {\em Mathematics of Operations Research}, 4(8), 1983.

\bibitem{MW}
A.~Meyerson and R.~Williams.
\newblock On the complexity of optimal k-anonymity.
\newblock In {\em PODS}, 2004.

\bibitem{Papa}
C.~Papadimitriou.
\newblock {\em Computational Complexity}.
\newblock Addison-Wesley, 1994.

\bibitem{SIGMOD}
H.~Park and K.~Shim.
\newblock Approximate algorithms for k-anonymity.
\newblock In {\em SIGMOD Conference}, 2007.

\bibitem{Sam-Swe}
P.~Samarati and L.~Sweeney.
\newblock Generalizing data to provide anonymity when disclosing information
  (abstract).
\newblock In {\em PODS}, 1998.

\bibitem{Swe02}
L.~Sweeney.
\newblock k-anonymity: a model for protecting privacy.
\newblock {\em Internation Journal on Uncertainity, Fuzziness and
  Knowledge-based Systems}, 10(5):557--570, 2002.

\end{thebibliography}

\end{document}